\documentstyle[seceq,epsf,wrapft]{ptptex}
\notypesetlogo  

\title{
Spatial resolution of near-field scanning optical microscopy with
sub-wavelength aperture}

\author{
Hiroaki {\sc Nakamura}\footnote{E-mail address: fdtd@harima.tcsc.nifs.ac.jp},
Keiji {\sc Sawada}$^1$, Hirotomo {\sc Kambe}$^1$, 
\\ Toshiharu {\sc Saiki}$^2$ and Tetsuya {\sc Sato}
}

\inst{
National Institute for Fusion Science

$^1$Department of Applied Physics, Faculty of Engineering, Shinshu University\\

$^2$Nano-Optical Dynamics Project, Kanagawa Academy of Science and Technology}

\recdate{
}

\abst{%
The finite-difference time-domain (FDTD) method is employed to solve the
three dimensional Maxwell equation for the situation of near-field
microscopy using a sub-wavelength aperture. Experimental result on
unexpected high spatial resolution is reproduced by our computer simulation.
}

\begin{document}

\newlength{\minitwocolumn}
\newlength{\minitwocolumna}
\newlength{\minitwocolumnb}
\setlength{\minitwocolumn}{0.5\textwidth}
\setlength{\minitwocolumna}{0.5\textwidth}
\setlength{\minitwocolumnb}{0.5\textwidth}
\addtolength{\minitwocolumn}{-2\columnsep}
\addtolength{\minitwocolumna}{+8\columnsep}
\addtolength{\minitwocolumnb}{-9\columnsep}

\maketitle

\section{Introduction}
Near-field scanning optical microscopy (NSOM)\cite{92b,98o} is a powerful
tool for the study of nanometer features with spatial resolution of 50-100
nm. The heart of NSOM is a near-field probe, which is a metal-coated optical
fiber tapered to sub-wavelength aperture. When the probe end approaches a
sample surface, the object is illuminated and the reemitted light is
collected in the near-field region of the aperture, whose diameter
determines the spatial resolution of NSOM. Optical imaging beyond the
diffraction limit is carried out by scanning the probe on the surface. In
addition to this fundamental principle, the resolution of NSOM is also
subject to the tapered structure of the probe. Such a behavior has been
demonstrated through our NSOM spectroscopy of single quantum dots.\cite{98s,99s}

Numerical analysis of electromagnetic field in the vicinity of the
aperture and propagation property of light in the tapered waveguide is quite
advantageous for the understanding of experimental results. We employ the
finite-difference time-domain (FDTD) method\cite{66y}
in the Mur absorbing boundary condition\cite{81m} to solve the
three-dimensional Maxwell equation for the same situation as the
experimental configuration and discuss the validity of simulation results.

\section{Calculations}
Figures 1 and 2 show the geometries of the problem. A near-field
fiber probe with a double tapered structure collects luminescence 
($\lambda=1\mu\mbox{m}$)
from a quantum dot buried $\lambda/10$ beneath the semiconductor surface. 
We assume
the source for luminescence is a point-like dipole current linearly
polarized along the $x$ direction. The radiation caught by $\lambda/2$ aperture is
transported to the tapered region clad with perfect conducting metal and
then coupled to the ordinary waveguide (optical fiber). We run the
simulation with time steps of ${\rm c \Delta} t=\lambda/ (40\sqrt{3})$ 
until the signal intensity
($|E_x|^2+|E_y|^2+|E_z|^2$) ,
which is evaluated at $(0, 0, 3.25\lambda),$ reaches steady state.

\noindent
\begin{minipage}[t]{\minitwocolumna}
\vspace*{0.3cm}
\section{Results}
\hspace{4ex} Figure 3 shows the calculated signal intensity as a function
of the displacement of the probe from the origin. For both scans along $x$
and
$y$ directions, the full width at half maximum of the signal  (spatial
resolution of NSOM) is estimated to be around $0.25\lambda,$
which is much smaller
than the aperture diameter of $\lambda/2.$ 
This performance is beyond the
fundamental principle of NSOM and in good agreement with the experimental
result.
Through  
\end{minipage}%
\hspace*{\columnsep}%
\begin{minipage}[t]{\minitwocolumnb}
\vspace*{-0.2cm}
\epsfxsize=\minitwocolumnb
\centerline{\epsfbox{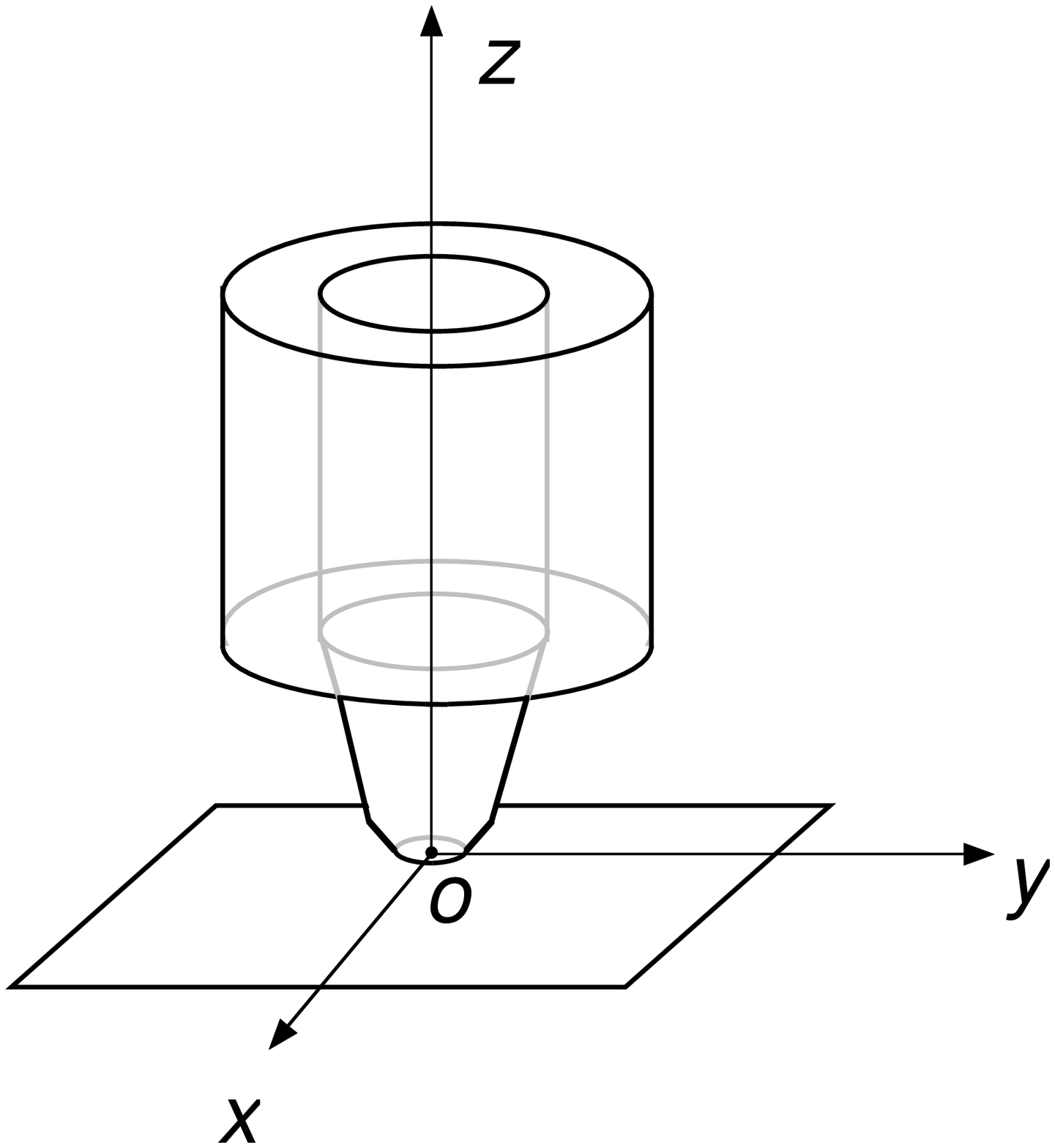}}
\refstepcounter{figure}
\label{fig.1}
\footnotesize
Fig. \thefigure.
Schematic picture of simulation for
NSOM.
\end{minipage}%
\vspace*{1ex}
this preliminary calculation, we demonstrate that 
FDTD simulation is quite useful to understand the behavior of light in the
near-field probe and to optimize its structure for advanced measurements.

\noindent
\begin{center}
\begin{minipage}[t]{\minitwocolumn}
\epsfysize=5.5cm
\centerline{\epsfbox{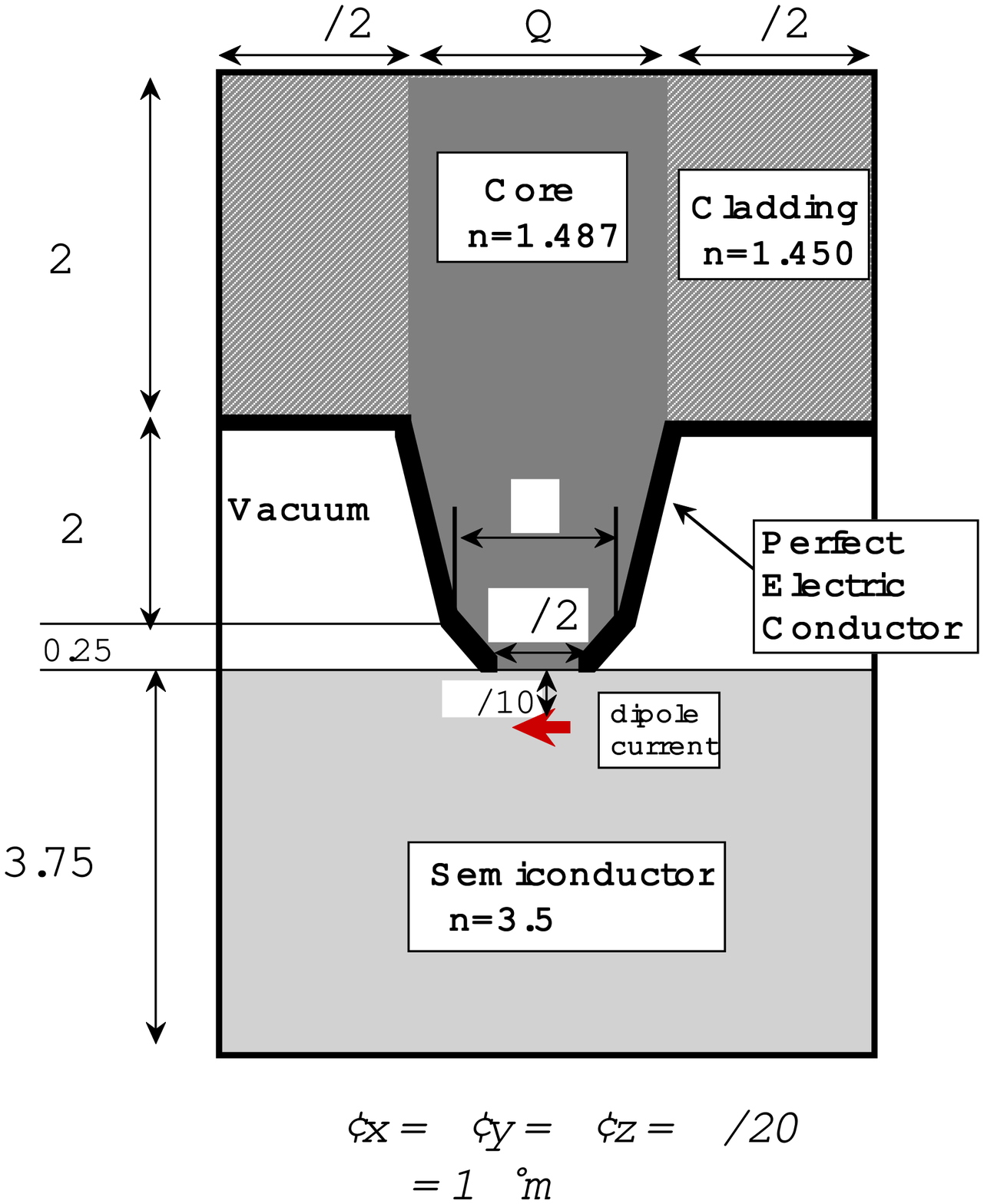}}
\refstepcounter{figure}
\label{fig.2}
\footnotesize
Fig. \thefigure.
Cross section diagram ($xz$-plane at $y=0$) of the geometry 
in our 3D computer simulations
for the double tapered fiber probe.
\end{minipage}%
\hspace*{\columnsep}%
\begin{minipage}[t]{\minitwocolumn}
\vspace*{-5.5cm}
\epsfysize=5cm
\centerline{\epsfbox{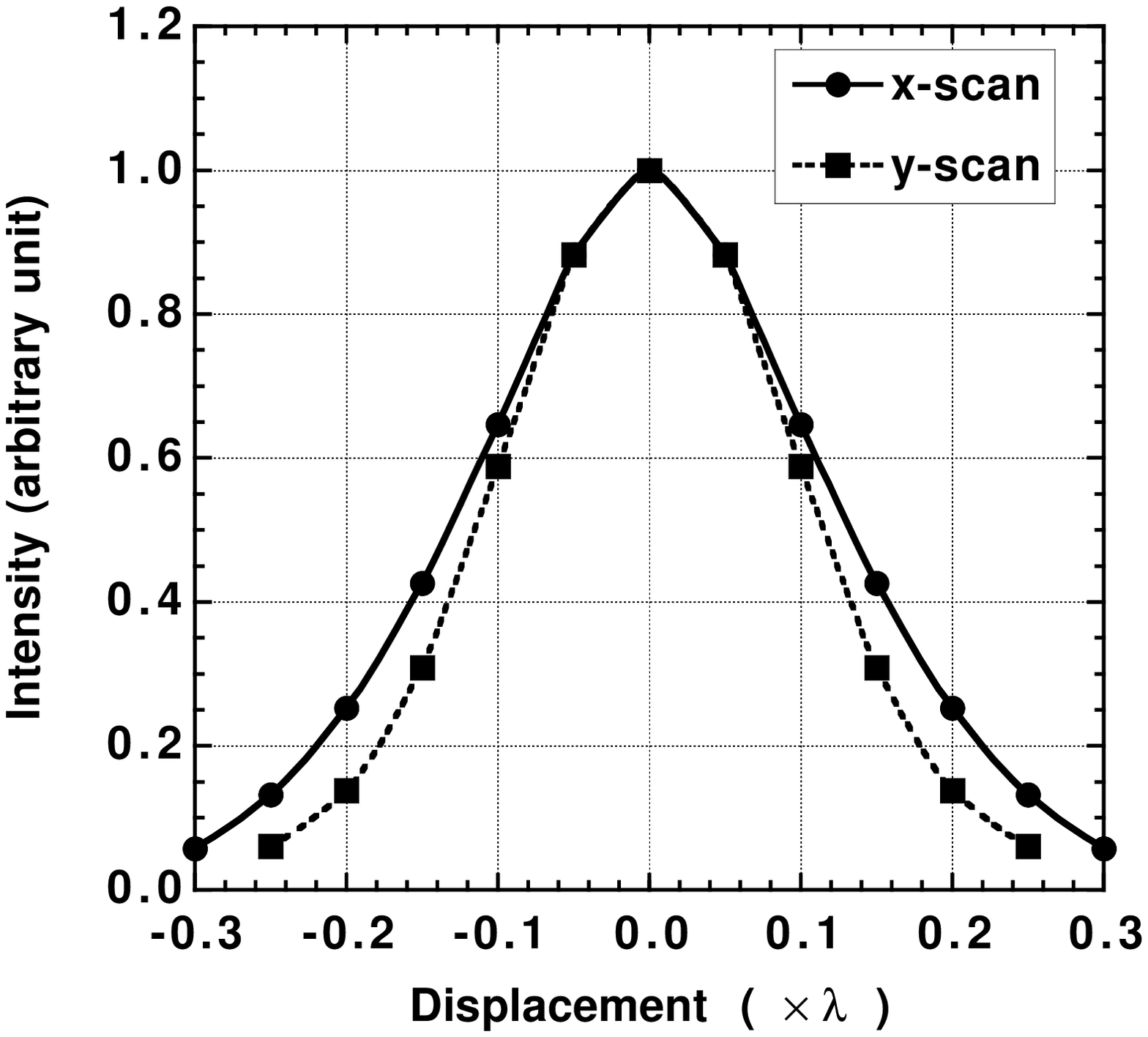}}
\refstepcounter{figure}
\label{fig.3}
\footnotesize
Fig. \thefigure.
Intensity vs. displacement from the origin $o$ in Fig.~\ref{fig.1}. 
Closed circles and squares denote the total electric field intensity
along $x$ and $y$ direction, respectively.
\end{minipage}%
\end{center}

\vspace*{2ex}
\noindent
This work was supported in part by the Grant-in-Aid
for Science Research from the Ministry of Education,
Science, Sports, and Culture. It was carried out by the Advanced Computing System
for Complexity Simulation (NEC SX-4/64M2) at National Institute
for Fusion Science.


\vspace*{-1ex}
\noindent
\begin{thebibliography}{99}
\bibitem{92b}
E.~Betzig and J.~K.~Trautman, \JL{Science, 257, 1992, 189}.
\bibitem{98o}
M. Ohtsu, ed., {\it Near-Field Nano/Atom Optics and Technology}
(Springer-Verlag, Tokyo, 1998).
\bibitem{98s}
T.~Saiki, K.~Nishi and M.~Ohtsu,
\JL{Jpn.~J.~Appl.~Phys., 37, 1998, 1638}.
\bibitem{99s}
T.~Saiki and K.~Matsuda, \JL{Appl.~Phys.~Lett., 74, 1999, 2773}.
\bibitem{66y}
K.~S.~Yee, \JL{IEEE Trans. Antennas Propag., AP-14, 1966, 302}.
\bibitem{81m}
G.~Mur, \JL{IEEE Trans. Electromagn. Compat., EMC-23, 1981, 377}.
\end{thebibliography}
\end{document}